\documentclass[a4paper]{article}
\usepackage[latin1]{inputenc} 
\usepackage[T1]{fontenc} 
\usepackage{RR,RRthemes}
\usepackage{hyperref}
\usepackage[ruled,commentsnumbered]{algorithm2e}
\RRdate{Decembre 2010}

\RRauthor{
Hien Thi Thu Truong 
  \and
Claudia-Lavinia Ignat 
}
\authorhead{Hien \& Claudia-Lavinia}
\RRtitle{Une approche reposant sur l'audit pour la gestion de la confiance dans la collaboration pair-à-pair }
\RRetitle{A Log Auditing Approach for \\ Trust Management in Peer-to-Peer Collaboration}
\titlehead{A Log Auditing Approach for Trust Management in Peer-to-Peer Collaboration}
\RRresume{Depuis quelques années nous assistons à une explosion de la
  popularité des logiciels sociaux comme les wikis, les blogs, les
  micro-blogs ou encore les réseaux sociaux tels que Facebook et
  MySpace. Malheureusement, les plus usités de ces services reposent
  tous sur un contrôle centralisé ; les données d'un utilisateur se
  retrouvant centralisées chez un seul fournisseur de service. Cela
  engendre indéniablement un problème de confidentialité puisque les
  utilisateurs n'ont plus le contrôle sur la manière dont leurs
  propres données sont diffusées. Afin de surpasser ces limitations,
  plusieurs approches récentes proposent de remplacer le contrôle
  centralisé de ces services par des approches pair-à-pair reposant
  sur des mécanismes de confiance où chaque usager décide avec qui il
  souhaite partager ces données personnelles. Toutefois, dans ce genre
  de collaboration, il est très difficile de s'assurer qu'une fois une
  donnée partagée avec des pairs, celle-ci ne soit pas divulguée à
  d'autres pairs non autorisés. Dans cet article, nous proposons un
  mécanisme qui permet de circonscrire les problèmes de violation de
  confidentialité liés à la divulgation de données par des pairs
  malicieux. Dans notre approche, les indices de confiance entre les
  usagers sont ajustés selon le comportement passé des usagers
  vis-à-vis des données partagées. Lorsqu'un usager partage des
  données, il impose aux autres usagers des obligations qu'ils doivent
  respecter. Chaque modification sur une donnée partagée effectuée par
  un usager et chaque obligation est répertoriée dans un journal. En
  réalisant un audit de ce journal, notre approche détecte les usagers
  qui se sont mal-comportés et ajuste en conséquence leurs indices de
  confiance.  } \RRabstract{Nowadays we are faced with an increasing
  popularity of social software including wikis, blogs, micro-blogs
  and online social networks such as Facebook and
  MySpace. Unfortunately, the mostly used social services are
  centralized and personal information is stored at a single
  vendor. This results in potential privacy problems as users do not
  have much control over how their private data is disseminated. To
  overcome this limitation, some recent approaches envisioned
  replacing the single authority centralization of services by a
  peer-to-peer trust-based approach where users can decide with whom
  they want to share their private data. In this peer-to-peer
  collaboration it is very difficult to ensure that after data is
  shared with other peers, these peers will not misbehave and violate
  data privacy. In this paper we propose a mechanism that addresses
  the issue of data privacy violation due to data disclosure to
  malicious peers. In our approach trust values between users are
  adjusted according to their previous activities on the shared
  data. Users share their private data by specifying some obligations
  the receivers must follow. We log modifications done by users on the
  shared data as well as the obligations that must be followed when
  data is shared.  By a log-auditing mechanism we detect users that
  misbehaved and we adjust their associated trust values by using any
  existing decentralized trust model.}
\RRmotcle{confiance, confidentialité des données, collaboration pair-à-pair,audit}
\RRkeyword{trust, privacy, peer-to-peer collaboration, log-auditing}
\RRprojets{SCORE}
\RRdomaine{2} 
 \RCNancy 

\begin{document}
\RRNo{7472}
\makeRR   
\section{INTRODUCTION}
\label{sec:introduction}

Social software including wikis, blogs, micro-blogs and social
networks has emerged as a new interpersonal communication form.
Existing micro-blogging services such as Twitter and social networks
such as Facebook or MySpace have millions of users using them
everyday. While these social services offer many attractive
functionalities, they require storing personal information in the
hands of a single large corporation which is a perceived privacy
threat. Users are obliged to provide and to store their data to
vendors of these services and to trust that they will preserve privacy
of their data, but they have little control over the usage of their
data after sharing it with other users.  These corporations could
produce a profile based on the individual behavior and therefore
detrimental decisions to an individual may be taken. Moreover, due to
large amounts of information these social services sites process every
day, a single flaw in the system could permit retrieval of large parts
of personal data. For instance on Facebook features such as messages,
invitations and photos help users gain access to private
information. Moreover, flaws in the Facebook's third-party API have
been found which allow for easy theft of private information.

Some recent approaches such as \cite{BucheggerEurosys09} proposed
moving away from centralized authority-based collaboration towards
a peer-to-peer trust network where users have full control over
their personal data that they store locally and can decide with
whom to share their data. Users define their network of trust
containing people that they trust and with whom they wish to
collaborate. These peer-to-peer networks of trust overcome the
disadvantages of centralized architectures by offering a good
scalability and fault-tolerance and the possibility of sharing
costs of administration. In a peer-to-peer collaboration model
rather than having a central authority which has access to all
users personal data, control over data is given to users.
Therefore, the risk of privacy breaches is decreased as well as
only a part of the protected data in the peer-to-peer network may
be exposed at any time. However, in this peer-to-peer
collaboration it is very difficult to ensure that after data is
shared with other peers, these peers will not misbehave and
violate data privacy. To prevent data misuse, trust management
mechanisms are used where peers are assigned trust values and a
peer collaborates only with high trusted peers. However, to our
knowledge, there exists no approach that automatically updates
trust values according to peers misbehavior.

In this paper we propose an approach of log auditing for computing
trust in a peer-to-peer environment according to respecting
obligations peers receive from other peers concerning their private
data. We also propose a novel audit-based compliance control
approach suited for distributed collaborative environments where
obligations are checked a-posterior and not enforced a-prior. This
approach in which usage policies are checked posteriorly is different
from prior-checked access control mechanisms. Rather than requiring a
hard security mechanism, our solution uses a trust-based approach that
is more flexible for users.

The rest of this paper is organized as follows. Section
\ref{sec:trust} is an overview of Peer-to-Peer trust. Section
\ref{sec:log-auditing} presents our log auditing approach in
decentralized systems. Then we describe the formal structure of
log in Section \ref{sec:log-structure}. In Section
\ref{sec:obligation} we present a discussion on obligations that
are associated to logs. Section \ref{sec:trust-assessment}
describes the mechanism for local trust assessment with algorithm
analysis. Section \ref{sec:related-work} compares our approach
with related works and Section \ref{sec:conclusion} presents
concluding remarks and directions of future work.

\section{PEER TO PEER TRUST OVERVIEW}
\label{sec:trust}

Peer-to-peer underlying architectures reflect society better than
other types of computer architectures \cite{Clark01}, being better
adapted to the way people think and to user's needs for knowledge
sharing and providing users more freedom to interact with each
other. In this peer-to-peer collaboration model, it is very difficult
to ensure data privacy. According to \cite{PrivacyWA67}, data privacy
is the right of individuals to determine for themselves \textit{when},
\textit{how}, and \textit{to what} extent information about them is
communicated to others. A peer shares his private data only with peers
that he trusts, so, privacy of data is preserved for a direct
connection with a peer. But the main issue refers to what happens to
data released to authorized persons, i.e how the user may, must and
must not use it. This issue is called usage control \cite{ParkTISS04,
  HiltyESORICS05} and is modeled by means of certain obligations that
users receive together with data.

Trust is a belief or confidence in the honesty, goodness of a person
or organization. In \cite{P2PComp09} a classification of trust models
is given. A trust level is an assessment of the probability that a
peer will not cheat. An honest peer will be assigned a high trust
level while a malicious peer will be assigned a low trust level. These
trust levels are updated according to the peer's behavior. If a peer
misbehaves, its trust level is decremented. The solution that we
propose in this paper for adjusting trust levels of peers according to
their behavior is general and could be combined with any existing
reputation mechanism.

In order to present an overview of our approach let us consider an
example in the domain of data sharing in a social network. Suppose
Alice creates a document and she wants to share it with different
friends, say Bob and Carol. She shares it to Bob with a certain
right to \textit{modify} it. It is very difficult to enforce Bob
to follow that policy in a decentralized environment. Bob can do
any action on the document once received it. There is no way for
Alice to guarantee Bob will not misbehave on the document after it
has been shared. In our approach we propose a mechanism logging
past actions of users concerning shared data. Bob's actions will
be logged by the system. Alice will never know what Bob has done
with her data if Bob only keeps the log locally. But his log of
local edit actions will be disclosed to whom he continues to
re-distribute data. If Carol receives the document, she can check
the log to know what actions that Alice and Bob did.

Our system does not aim to prevent fraud. Rather, the log mechanism
provides audit capabilities in order to detect attempts at fraud after
the data has been shared and used. The local actions on data and the
communications between peers are assumed to leave some evidence and
hence are observable. The owner attaches a usage policy to the data in
order to specify what actions are allowed and under which
conditions. According to this a-posterior checking, user trust values
are updated with a decrement. For checking compliance to obligations,
we audit the log containing modifications done by users on the shared
data as well as the obligations that must be followed. Each user
evaluates trust on other users and keeps trust values locally instead
of storing them at a central authority. Trust values help users decide
to collaborate or not with other users. These values are updated after
each posterior checking of the log for detection of misbehaved
users. We can use any trust model for updating trust values. To our
knowledge, our approach is the only one that addresses data privacy
violation by discovering malicious users and updating user trust
values.

\section{LOG AUDITING APPROACH}
\label{sec:log-auditing}

Our system consists of a group of communicating peers. Each peer has
its own workspace. These peers collaborate together in creating and
sharing data in a decentralized environment where no central
administration point exists. Users are administrators themselves.

The local edit actions and communication actions among peers are
logged by the system in \textit{edit log} and \textit{communication
  log}. Each user keeps locally one edit log and one communication
log. When a user shares the document with others, logs and usage
policies will be associated with the document. The policies are
specified in communication log. Initially, the log is empty, but after
certain iterations, as observations are made, the log will grow
up. The logs are created under the following assumptions:

\begin{itemize}

\item[$\diamond$]Logs are created automatically by system and they are
  unalterable. This assumption is practical. In reality, logs could be
  changed but there are some techniques as in \cite{BrentCDCS03} to
  detect or avoid log modification.
\item[$\diamond$] It is required that at least one obligation is given
  in a sharing action. In collaboration, when the document is sent
  back to previous sender, it is not required to re-define new
  obligations.
\item[$\diamond$] The occurring order of events stored in logs is
  maintained by logical clock \cite{TimeLL78}.
\end{itemize}

In a collaboration-based system, users are expected to behave
correctly, but they might be suspected of incorrect behaviors. A user
violates an obligation if he performs actions which are not permitted
in usage policies. We update decremently their trust values each time
violations are detected. The trust value for well-behaved peers is
higher than the trust value for malicious peers. A peer has initial
trust values associated with other peers. They are calculated and
adjusted after each log analysis.

Log auditing consists in the analysis of both edit and
communication logs.  Each time the user receives different
versions of a same document, the system automatically analyzes the
logs in order to judge the past behaviors of other users. An
important point in checking past behaviors is to detect
mismatching of actions and obligations them. Next, the received
logs that include past actions and obligations are checked to be
merged with current local logs. If the logs should be merged, both
document edit logs and communication logs are merged. In addition
to merging obligations, conflict resolution between rights is
required.
\section{LOG STRUCTURE}
\label{sec:log-structure}
In this section, we present the structure of logs and give an example
to illustrate how logs are created and stored locally at sites of
peers.

\textbf{Definition 1.} \emph{An event is denoted as $e = a_{r} ^{s}$
  where $a$ represents an action that can be either a local action on
  document or a communication action, and $r$ represents parameters
  which are in form of pair of name and value.  The notation $s$ has
  the temporal meaning for an event, in which $s = -1$ represents the
  actual event that a user performed while $ s = +1$ represents the
  obligation event he has to follow.}

Similar to event structure in Z language \cite{PretschnerSTM08}, each
event in our log is composed of an action or an obligation and several
parameters. For instance, $share ^ {-1}_{\{by, P_{1}\},\{to,P_{2}\}}$
is an event of sharing the document from $P_{1}$ to
$P_{2}$ ($P_{1}$ shared document to $P_{2}$) while
$share^{+1}_{\{by,P_{1}\},\{to,P_{2}\}}$ is an obligation $P_{1}$
gives to $P_{2}$ ($P_{2}$ can share the document).

\textbf{Definition 2.} \emph{A log is composed of a time-ordered
  sequence of pairs (logical clock, event): $[(c_{1}$,
    $e_{1})$, $(c_{2}$, $e_{2})$,..., $(c_{m}$, $e_{m})]$.}

  For ordering events we use the logical clock with
  \emph{happened\_before} relation \cite{TimeLL78} among events. Event
  $e_{1}$ is ordered before event $e_{2}$ if $e_{1}$ happened before
  $e_{2}$. The system of each site maintains a counter that is
  incremented each time an event is generated at that site. Each event
  is assigned the value of the counter at the moment of its
  generation. This counter called also logical time $c_{j}$ is simply
  used to order events according to their order of occurrence. The
  logical clock of obligations of sender is replaced by the new
  logical clock of receiver according to the order when he receives
  the document. This helps checking if events generated by a user
  conform to obligations previously received. We can track backward
  the logical clock value assigned to obligation by sender through the
  logical clock value of \textit{share} event. From the logical clock
  of \textit{share} action, we know when such obligation events are
  shared by the sender.

\begin{figure}[ht]
\centering
    \includegraphics[scale=0.24]{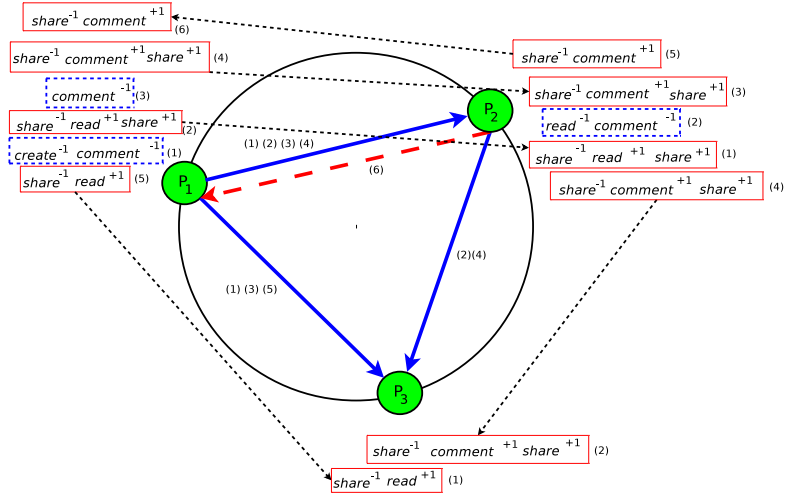}
    \caption{Logs for 3 peers collaboration, \textit{edit} actions
      are inside dash box and \textit{communication} actions are
      inside line-box. Only local edit actions of each peer and
      obligations are showed in this figure.}
\end{figure}

Consider an example of sharing data in a distributed peer to peer
social network. Users can \emph{share} photos, videos or music
documents between them and add \emph{comment} to these documents (but
they can not modify the content of these documents). The obligations
could be: \emph{``may read''}, \emph{``may not read''}, \emph{``may
  add comment''}, \emph{``may not add comment''}, \emph{``may
  delete comment''}, \emph{``may not delete comment''}, \emph{``may
  share''}, \emph{``can not be shared''}.
Figure 1 shows an example of three peers $P_{1}$, $P_{2}$, $P_{3}$
sharing photos.  Let $P_{1}$ be the creator of data $d$. $P_{1}$
shares $d$ with $P_{2}$, and then $P_{2}$ shares it with $P_{3}$.
In parallel, $P_{1}$ shares the same data directly to $P_{3}$. The
logs of actions (edit log $l_{P_{x}-edt}$ and communication log
$l_{P_{x}-com}$) are created locally at peers as follows:
\begin{enumerate}
\item At the local site, $P_{1}$ creates document $d$ for which he
adds a comment and shares this document with
  $P_{2}$ with the usage obligation \emph{`` may read''}, \emph{``may
    share further''}.  Logical clock starts from $1$.

$\bf l_{P_{1}-edt, d}$ = $[(1, create_{\{by, P_{1}\}}^{-1})$, $(1,comment_{\{by,P_{1}\}}^{-1})]$;

$\bf l_{P_{1}-com, d}$ (with $P_{2}$) = $[(2,share_{\{by,P_{1}\}, \{to, P_{2}\}}^{-1})$,

$(2,read_{\{by,P_{1}\}, \{to, P_{2}\}}^{+1})$, $(2,share_{\{by,P_{1}\}, \{to, P_{2}\}}^{+1})$,

$(2,not \ comment_{\{by,P_{1}\}, \{to, P_{2}\}}^{+1})]$ \\

\item $P_{2}$ receives document $d$ with associated logs and adds a
  comment to this document. The edit log will be updated
  continuously. Note that the logical clock of obligations in
  communication log is updated to the value of the local logical clock
  at the site $P_{2}$. Note also that $P_{2}$ received document $d$
  with the permission of reading it and sharing it further, but
  without the permission of commenting it. However he did the comment
  action, therefore, $P_{2}$ violated the received obligation.

  $\bf l_{P_{2}-com,d}$ = $[(2,share_{\{by,P_{1}\}, \{to,
    P_{2}\}}^{-1})$, $(1,read_{\{by,P_{1}\}, \{to, P_{2}\}}^{+1})$,
  
  $(1,share_{\{by,P_{1}\}, \{to, P_{2}\}}^{+1})$, $(1,not \
  comment_{\{by,P_{1}\}, \{to, P_{2}\}}^{+1})]$ \\

  $\bf l_{P_{2}-edt,d}$ = $[(1, create_{\{by, P_{1}\}}^{-1})$,
  $(1,comment_{\{by,P_{1}\}}^{-1})]$ $\cup$

  $[(2,read_{\{by,P_{2}\}}^{-1})$, $(2,comment_{\{by,P_{2}\}}^{-1})]$;

\item After sending to $P_{2}$, $P_{1}$ adds another comment to
  document $d$ and re-sends to $P_{2}$ this document with obligation
  \emph{``may add comment''}. The logical clock of $P_{1}$ is
  increased after each action.  $P_{1}$ also shares data with $P_{3}$
  with the same obligation \emph{``may add comment''}.

  $\bf l_{P_{1}-edt,d}$ = $[(1, create_{\{by, P_{1}\}}^{-1}$,
  $(1,comment_{\{by,P_{1}\}}^{-1})$,

  $(3,comment_{\{by, P_{1}\}}^{-1})]$;

  $\bf l_{P_{1}-com,d}$ (with $P_{2}$) = $[(2,share_{\{by,P_{1}\},
    \{to, P_{2}\}}^{-1})$,

  $(2,read_{\{by,P_{1}\}, \{to, P_{2}\}}^{+1})$,
  $(2,share_{\{by,P_{1}\}, \{to, P_{2}\}}^{+1})$,

  $(2,not \ comment_{\{by,P_{1}\}, \{to, P_{2}\}}^{+1})$,

  $(4,share_{\{by,P_{1}\}, \{to, P_{2}\}}^{-1})$,
  $(4,comment_{\{by,P_{1}\}, \{to,P_{2}\}}^{+1})]$

  $\bf l_{P_{1}-com,d}$ (with $P_{3}$) = $[(5,share_{\{by,P_{1}\},
    \{to,P_{3}\}}^{-1})$,

  $(5,comment_{\{by,P_{1}\}, \{to, P_{3})\}}^{+1})]$.

\item $P_{2}$ receives document $d$ again from $P_{1}$ and then shares
  it with $P_{3}$ with the obligation \emph{``not share''} to do not
  share further the document. The edit log of $P_{2}$ is updated to
  include the last action done by $P_{1}$ of commenting the document.

  $\bf l_{P_{2}-com,d}$ (local) = $[(2,share_{\{by,P_{1}\}, \{to,
    P_{2}\}}^{-1})$,

$(1,read_{\{by,P_{1}\}, \{to, P_{2}\}}^{+1})$,
$(1,share_{\{by,P_{1}\}, \{to, P_{2}\}}^{+1})$,

$(1,not \ comment_{\{by,P_{1}\}, \{to, P_{2}\}}^{+1})]$

$\cup$ $[(4,share_{\{by,P_{1}\}, \{to, P_{2}\}}^{-1})$,
$(3,comment_{\{by,P_{1}\}, \{to,P_{2}\}}^{+1})$

$\bf l_{P_{2}-edt,d}$ = $[1, create_{\{by, P_{1}\}}^{-1}$,
$(1,comment_{\{by,P_{1}\}}^{-1})]$

$\cup$ $[(2,read_{\{by,P_{2}\}}^{-1})$,
$(2,comment_{\{by,P_{2}\}}^{-1})]$ $\cup$

$[(3,comment_{\{by, P_{1}\}}^{-1})]$

$\bf l_{P_{2}-com,d}$ (with $P_{3}$) = $\bf l_{P_{2}-com,d}$ (local)
$\cup$

$[(4,share_{\{by, P_{2}\}, \{to, P_{3}\}}^{-1})$,$(4,not \
share_{\{by,P_{2}\}, \{to, P_{3}\}}^{+1})]$.

\item At local site of $P_{3}$, suppose that $P_{3}$ receives
document $d$ from $P_{1}$ before receiving it from $P_{2}$. The
local communication log of $P_{3}$ is obtained by merging $\bf
l_{P_{1}-com,d}$ (with $P_{3}$) computed at step 3 with $\bf
l_{P_{2}-com,d}$ (with $P_{3}$) computed at step 4.

$\bf l_{P_{3}-com,d}$ = $[(5,share_{\{by,P_{1}\},
  \{to,P_{3}\}}^{-1})$,

$(1,comment_{\{by,P_{1}\}, \{to, P_{3}\}}^{+1})]$ $\cup$
$l_{P_{2}-com}$ (local)

$\cup$ $[(4,share_{\{by, P_{2}\}, \{to, P_{3}\}}^{-1})$, $(2, not \
share_{\{by,P_{2}\}, \{to, P_{3}\}}^{+1})]$

$\bf l_{P_{3}-edt,d}$ = $\bf l_{P_{2}-edt,d}$

\end{enumerate}

In order to detect cheaters, each peer analyzes the received logs. A
user with actions that do not conform to obligations is considered as
a cheater. In the above example, $P_{3}$ detects the violation of
action \textit{comment} of $P_{2}$ in $l_{P_{2}-edt}$.
\section{OBLIGATIONS}
\label{sec:obligation}

Collaborating in a distributed system makes a user possible to receive
document from many collaborators. In the previous example $P_{3}$
receives the same document from $P_{1}$ and $P_{2}$. $P_{3}$ will
update document based on the received changes under certain
obligations. Up on the obligations, the system decides to accept or
reject the new copies of data.

When a user receives different obligations, the conflict between
obligations may occur. An obligation conflict means the subjects
are both required and required not to perform the same actions on
target objects. In multi-policies environment, it is possible that
one policy overrides another. Conflict detection should be
performed in order to decide which usage policy is performed and
which is ignored. Moreover, in case a user receives a set of
obligations instead of a single obligation, the conflict may occur
between sets. Two sets are in conflict if they contain at least
one conflict between two single obligations. One of the solution
to avoid conflict is giving priority to certain obligations.

Obligations can be ordered based on the ability they offer to work on
data. We denote two obligations $\alpha_{1}$, $\alpha_{2}$ that
$\alpha_{1}> \alpha_{2}$ if $\alpha_{1}$ has more ability to work on
document than $\alpha_{2}$. For example $add-comment> read$ and
$share > not \ share$. The comparison of two sets of obligations can
be performed based on comparing each single obligation belonging to
the two sets. For example, $[add-comment$, $share]$ $>$ $[not \
add-comment$, $share]$.

With the approach using obligations in sharing document, if some
obligation is not specified, for example, neither $share$ nor $not \
share$ is given, peers can do the actions as they want without any
violation, e.g they can either $share$ or $not \ share$ the document.

In obligation-based collaborating systems, the more user respects
obligations, the more trust he gains, and the more possibility others
want to collaborate with him.

Unlike single centralized system, in distributed P2P application,
peers are faced with conflicts between rights and obligations
referring to the same document, but also between changes on the same
document. When a peer receives many copies of the same document, it
analyzes the associated logs in order to assess the local trust values
of other peers who collaborated on copies.  Afterward that peer checks
for merging the logs and the document. If the obligations permit that
peer to get the changes on document, a merge algorithm is
performed. Due to space limitation, we do not present in this paper
our algorithms related to merging document and ordering rights and
obligations.
\section{TRUST ASSESSMENT}
\label{sec:trust-assessment}

\begin{algorithm}[t]
  \footnotesize{ \small{ \SetKwIF{If}{ElseIf}{Else}{if}{then}{else
        if}{else}{end if} \SetKwFor{For}{for}{do}{end for}
      \SetKwFor{While}{while}{do}{end while} \KwIn{The edit log
        $l_{edt}$ and the communication log $l_{com}$, the document
        $d$, user $A$ who assessing trust.}
      \KwOut{$T_{A}^{log}(P_{i})$ for each $P_{i}$ that appears in logs.}
\Begin
{
    \For {each $(c,e=(a)_r^{s})$ with $s=-1$}
        {
            $misbehaved$ = \textit{FALSE}\;
            extract $a$, $P_{i}$ in $\{by,P_{i}\}$ from $r$\;
            $T_{A}^{log}(P_{i})=max\_trust\_value$\;
            \If {$P_{i}$ is not the creator of d}
            {
                $k =lengthOf(l_{com})$\;
                $checked$ = \textit{FALSE}\;
                \While {($k \ge 1$) and ($checked$ = \textit{FALSE})}
                {
                    get $(c_{k},e_{k} = (b)_{r}^{s}) \in l_{com}$ \;
                    extract $b$, $P_{j}$ in $\{to, P_{j}\}$ from $e_{k}$ \;
                    \If {$(s=+1)$ and $(P_{j}= P_{i})$ and $(c_{k} < c)$}
                    {
                        \If {($b = not\  a$)}
                        {
                            $misbehaved$ = \textit{TRUE}\;
                            $checked$ = \textit{TRUE}\;
                        }
                    }
                    $k=k-1$\;
                }
                \If {($misbehaved$ = \textit{TRUE})}
                    {
                        adjust decemently trust value $T^{log}_{A}(P_{i})$ based one specific trust model;
                    }
            }
        }
\caption{LOCAL-TRUST-ASSESSMENT}\label{trust-assessment}
}
}
}
\end{algorithm}

In our decentralized system, each peer evaluates trustworthiness of
other peers based on its experience. During collaboration between
peers, trust values are adjusted mainly upon the result of log
analysis. Checking a log is a basic mechanism to detect cheaters and
help to predict the probability that they will continue cheat in
future actions.

We denote $T^{log}_{P_{i}}(P_{j})$ as the trust value that a peer
$P_{i}$ evaluates and assigns to peer $P_{j}$. In order to manage
trust values for peer $P_{j}$, we can use any existing
decentralized trust model. The trust values are initially assigned
a default value by system.

The algorithm \ref{trust-assessment} takes as input linear logs
(edit log and communication log). This is a local algorithm that
peers can apply in order to determine trust on other peers over
the collaborative network. The peer $P_{i}$ updates value
$T^{log}_{P_{i}}(P_{j})$ for peer $P_{j}$ based on the result of
log analysis.

All peers are set the highest trust value at the beginning
($max\_trust\_value$). In order to detect misbehaviors, peer's actions
are considered violating the obligations if there is one right or
obligation which not permit to do that action.  With each event in
log, parameter $(by,P_{i})$ helps extracting the user $P_{i}$ who
performed the action. We consider action is made by $P_{i}$ at logical
time $c$. If $P_{i}$ is the creator of the document, it has full
rights to do any action on the document, therefore, no need to check
for its obligation. In case $P_{i}$ has received the document from
another peer, we will check for its actions to compare with the given
obligations. The obligations are kept as a special ``event'' in
communication log ($c,(b)^s_{r}$) with $s=+1$. We extract parameter
($to,P_{j}$) from this ``event''. If $P_{j} = P_{i}$ and the logical
clock value $c_{j}$ when obligation was received is less than logical
clock value $c_{i}$ when action was performed, that action is
considered valid. As the logical clock of obligation is transformed
from sender's to receiver's, we can check whether an action was done
before or after a peer received the corresponding obligation. It
should be noticed that rights or obligations are possible to be
overridden and the latest ones are taken in account in our algorithm
only.

When an  assessed peer $P_{j}$ is detected as a cheater, its local
trust value is decremented by assessing peer $P_{i}$. The local
trust values could be aggregated from log-based trust, reputation
or recommendation trust. That depends on the trust model being
used. Research on the trust models is out of scope of this paper.

Our algorithm serves for trust assessment by using logs. The violation
in case a cheating user copies the content of document to create a new
one, then claims him as owner can not be detected by using log
auditing itself. However, communication log could be used to discover
the history actions on document that helps to detect cheaters.

\section{Related Work}
\label{sec:related-work}

In this section we first compare our work with some approaches
that address data privacy in peer-to-peer systems. Then we
continue by describing and comparing our proposed mechanism with
other approaches that use some related solutions to our approach
but in different contexts.

In order to return data ownership to users rather than to a third
party central authority, some recent works
\cite{BucheggerEurosys09,WolinskyCOPS10} explore the coupling between
social networks and peer-to-peer systems. In this context privacy
protection is understood as allowing users to encrypt their data and
control access by appropriate key sharing and distribution. Our
approach is complementary to this work and refers to what happens to
data after it has been shared.

Another approach that addresses data privacy violation in
peer-to-peer environments is Priserv \cite{JawadGlobe09}, a DHT
privacy service that combines the Hippocratic database principles
with the trust notions. Hippocratic databases enforce
purpose-based privacy while reputation techniques guarantee trust
notions. However, this approach focuses on a database solution,
being limited to relational tables. Moreover, as opposed to our
solution, the Priserv approach does not propose neither a
mechanism of discovering the malicious users that do not respect
the obligations required for using the data nor an approach for
updating the trust values associated to users.

OECD (Organization for Economic Cooperation and Development)
defined basic privacy principles including: collection limitation,
data quality, purpose specification, use limitation, security
safe, openness, individual participation, accountability. We
consider data privacy in collaborative working from the point of
\textit{use limitation} that users will specify how their data may
and may not be used. We consider the decentralized system which
documents are exchanged and shared among users. When a user
receives a document, he is expected to work on the document by
respecting obligations. The log mechanism is used to detect
cheaters who do not respect their obligations. Unlike access
control which is concerned with granting access to sensitive data
based on conditions that relate to past or present, obligation
which impose conditions on the future is concerned with
commitments of the involved users. At the moment access is granted
to data, adherence to these commitments cannot be ensured. The
formal framework in \cite {PretschnerESORICS05} allows
specification of obligations. They present different mechanisms
for checking adherence to commitments. However, all their proposed
solutions are based on a central reference monitor that can ensure
that data protection requirements are adhered to. As opposed to
our approach, these solutions are not suitable for peer-to-peer
environments where there is no central authority.

Keeping and managing event logs is frequently used for ensuring
security and privacy. This approach has been studied in many
works. In \cite{CederquistIS09}, a log auditing approach is used
for detecting misbehavior in collaborative work environments,
where a small group of users share a large number of documents and
policies. In \cite{KarlCS08,RogerCSFW01}, a logical policy-centric
for behavior-based decision-making is presented. The framework
consists of a formal model of past behaviors of principals which
is based on event structures. However, these models require a
central authority to audit the log to help the system making
decisions and this is a limitation for using these models in a
fully decentralized environment.

Trust management is an important aspect of the solution that we
proposed. The concept of trust in different communities varies
according to how it is computed and used. Our work relies on the
concept of trust which is based on past encounters: ``Trust is a
subjective expectation an agent has about another's future behavior
based on the history of their encounters'' \cite {MuiHICSS02}. Various
trust models for peer to peer systems exist such as NICE model
\cite{LeeCT06}, EigenTrust model \cite{EigenTrustWWW03} and global
trust model \cite{GlobalTrustKM01} and our mechanism for discovering
misbehaved users can be coupled with any existing trust model in order
to manage user trust values.


\section{Conclusion}
\label{sec:conclusion}
Our vision is to replace central authority-based social software
collaboration with a distributed collaboration that offers support for
decentralization of services. In this context, our paper addressed the
issue of data privacy violation due to data disclosure to malicious
peers in a peer-to-peer collaboration.  In our collaboration model
users share their private data by specifying some obligations the
receivers must follow. Modifications done by users on the shared data
and the obligations that must be followed when data is shared are
logged in a distributed manner. A mechanism of distributed log
auditing is applied during collaboration and users that did not
conform to the required obligations are detected and therefore their
trust value is updated. Any distributed trust model can be applied to
our proposed mechanism. Users can perform concurrent modifications on
the shared documents as well as they can share documents with
different specified obligations according to their preferences.


A direction of future work is the evaluation of the proposed
mechanism.  We will test first the efficiency and complexity of
our algorithms in peer-to-peer simulators such as PeerSim
\cite{peersim}. We plan afterward to apply our trust management
approach to existing research peer-to-peer online social networks such
as PeerSoN \cite{peerson}.

%
%
\bibliographystyle{abbrv}
\bibliography{ref}

\end{document}